\documentclass[journal]{IEEEtran}
\usepackage{tensor}
\usepackage{graphicx}
\usepackage{subcaption}
\usepackage{tabularx}
\usepackage[cmex10]{amsmath}
\usepackage{amsthm}
\usepackage{amssymb}
\usepackage{bm}
\usepackage{algorithm}
\usepackage{algorithmic}
\usepackage{cite}
\usepackage{setspace}
\usepackage{stfloats}
\usepackage{enumerate}
\usepackage{cases}
\usepackage{multirow}
\usepackage{mathrsfs}
\usepackage{array}
\usepackage{color}
\usepackage{verbatim}
\usepackage{extpfeil} 
\usepackage{booktabs}
\usepackage{multirow}
\usepackage{caption}
\usepackage{graphicx}
\usepackage{tabularx}
\usepackage{epstopdf}
\usepackage{textcomp}
\usepackage{epsfig,epsf,color,balance,cite}
\usepackage{amssymb}
\usepackage{amsthm}
\usepackage{graphicx}
\usepackage{array,color}
\usepackage{amsmath}
\usepackage{graphicx}
\usepackage{tabularx}
\usepackage{epsfig,epsf,color,balance,cite}
\usepackage{algorithmic}
\usepackage{algorithm}
\usepackage{bm}
\usepackage{caption}
\usepackage{textcomp}
\usepackage{color}
\usepackage{multirow}
\usepackage{cite}
\usepackage{enumerate}
\usepackage{cases}
\usepackage{color}
\usepackage{epstopdf}
\usepackage{textcomp}
\usepackage{url}
\usepackage[colorlinks]{hyperref}

\def\BibTeX{{\rm B\kern-.05em{\sc i\kern-.025em b}\kern-.08em
    T\kern-.1667em\lower.7ex\hbox{E}\kern-.125emX}}
\begin{document}

	\makeatletter
	\newcommand{\rmnum}[1]{\romannumeral #1}
	\newcommand{\Rmnum}[1]{\expandafter \@slowromancap \romannumeral #1@}
	\makeatother
	
	\title{Near-Field Mobile Tracking: A Framework of Using XL-RIS Information}
	
	\author{  Tuo Wu, Cunhua Pan,  Kangda Zhi, Junteng Yao, Hong Ren,   Maged Elkashlan, and Chau Yuen, \emph{Fellow, IEEE}
\thanks{\emph{(Corresponding author: Cunhua Pan and Chau Yuen.)}}
\thanks{T. Wu and C. Yuen are with the School of Electrical and Electronic Engineering, Nanyang Technological University, 639798, Singapore (E-mail: $\rm \{tuo.wu, chau.yuen\}@ntu.edu.sg$). C. Pan, H. Ren are with the National Mobile Communications Research Laboratory, Southeast University, Nanjing 210096, China. (E-mail: $\rm \{cpan, hren \}@seu.edu.cn$). K. Zhi is with the School of Electrical Engineering and Computer Science, Technical University of Berlin, 10623 Berlin (E-mail: $\rm k.zhi@tu\text{-}berlin.de$). J. Yao is with the Faculty of Electrical Engineering and Computer Science, Ningbo University, Ningbo 315211, China (E-mail: $\rm  yaojunteng@nbu.edu.cn$).  M. Elkashlan is with the School of Electronic Engineering and Computer Science at Queen Mary University of London, London E1 4NS, U.K. (E-mail: $\rm maged.elkashlan@qmul.ac.uk$).
  }
}
	
	\markboth{}
	{}
	\maketitle

	\begin{abstract}
This paper introduces a novel mobile tracking framework leveraging the high-dimensional signal received from extremely large-scale (XL) reconfigurable intelligent surfaces (RIS). This received signal, named XL-RIS information, has a much larger data dimension and therefore offers a richer feature set compared to the traditional base station (BS) received signal, i.e., BS information, enabling more accurate tracking of mobile users (MUs).   As the first step,  we present  an XL-RIS information reconstruction (XL-RIS-IR) algorithm to reconstruct the high-dimensional  XL-RIS information from the low-dimensional BS information.  Building on this, this paper  proposes a comprehensive framework for mobile tracking, consisting of a \emph{Feature Extraction Module} and a \emph{Mobile Tracking Module}. The \emph{Feature Extraction Module} incorporates a convolutional neural network (CNN) extractor for spatial features, a time and frequency (T$\&$F) extractor for domain features, and a near-field angles of arrival (AoAs) extractor for capturing AoA features within the XL-RIS.  These features are combined into a comprehensive feature vector, forming a time-varying sequence fed into  the  \emph{Mobile Tracking Module}, which employs an Auto-encoder (AE) with  a stacked bidirectional long short-term memory (Bi-LSTM) encoder and a standard LSTM decoder to predict MUs' positions in the upcoming time slot. Simulation results confirm that the tracking accuracy of our proposed framework is significantly enhanced by using reconstructed XL-RIS information and exhibits substantial robustness to signal-to-noise ratio (SNR) variations.
	\end{abstract}

	\IEEEpeerreviewmaketitle

\section{Introduction}
The upcoming introduction of sixth generation (6G) Internet of Things (IoT) wireless networks marks a pivotal moment in our technological journey \cite{Wang1, ChenHui1, Wang3, FAS1, FAS2}. Enhanced localization accuracy becomes central in this new era  \cite{ChenHui2}.  The heightened demand for accuracy challenges us to not only innovate but to fundamentally rethink our approach to localization \cite{Kaitao, HenkWymeersch1, HeJiguang1, Wang4}.  Recently, reconfigurable intelligent surfaces (RIS) \cite{Wang2, Cunhua1, Zhi, Gui1, Zhi3} have been introduced as a revolutionary technology to enhance the localization accuracy \cite{Wu1,JiguangHe2, Wang5} with several advantages. First, RIS offers a cost-effective and energy-efficient alternative to traditional base stations (BS) for maintaining consistent connectivity, especially in challenging environments. Additionally, their slim and adaptable design easily deploys in the urban structures, making them ideal for 6G networks.

RIS-aided localization algorithms can be categorized into two main types:  \emph{two-step method} \cite{chen2023multirisenabled} and \emph{fingerprint-based} \emph{method} \cite{WuFingerprint}. The \emph{two-step method} estimates parameters like angle of arrival (AoA) and time  of arrival (ToA) to deduce the location by exploiting geometric relationships \cite{ye2017power, hu2020deep, huang2018deep, yang2019deep}. The \emph{fingerprint-based method} \cite{ Bhattacherjee, 3DCNN}, on the other hand, relies on a database of pre-recorded signal characteristics, such as received signal strength (RSS), and compares real-time signals to the database for pinpointing a location. Building upon these two methods, algorithm design for RIS-aided localization has been widely investigated,  focusing on various applications and different perspective, such as multi-user \cite{Kamran2}, multi-RIS \cite{Wu2RIS}, and sidelink scenarios \cite{chen2023multirisenabled}.

It is worth noting that the above studies were built based on the far-field channel model, with the help of tractable planar wavefront assumptions.  To overcome the high path-loss attenuation and fully unleash the gain of RIS,  the RIS is expected to be constructed on an extremely large scale and equipped with large numbers of passive and low-cost reflecting elements. However, with a large physical dimension, the communication between an RIS and users will be located in the near field, rather than the conventional far field \cite{WangR, 2012Bias, HenkWymeersch1}. Recognizing this, recent research \cite{Shubo,Xing,Mingan,Chongwen1,Cuneyd,Reza} has significantly advanced the design of RIS-aided localization algorithms in the presence of the near-field effect. These studies have introduced various methodologies to enhance localization accuracy, from creating virtual line-of-sight (VLoS) paths using RIS \cite{Shubo} to employing advanced optimization techniques for precision improvement \cite{Xing}. Efforts have also been made to explore the theoretical limits of holographic RIS-aided localization system \cite{Chongwen1} and address specific challenges such as phase-dependent amplitude variations \cite{Cuneyd} and bi-static sensing for fixed transmitter and receiver systems \cite{Reza}.

Besides the research limitation concerning the near field, another important challenge is about mobile localization aiming to accommodate the movement of mobile user (MU). While the static approaches developed in earlier works \cite{Shubo,Xing,Mingan,Chongwen1,Cuneyd,Reza} laid a strong foundation, their applicability is somewhat constrained in scenarios featuring mobile users. Addressing this need, more recent research \cite{Palmucci, NF2, NF3}  have switched the focus towards designing the mobile tracking algorithms. Specifically, the authors in \cite{Palmucci} delved into jointly optimizing the RIS reflection coefficients and BS precoding to adeptly track MUs in scenarios with multi-RIS setups. Additionally, recognizing challenge that integrated the mobility and LoS obstructions, Mei  \emph{et al.} in \cite{NF2}  developed a multi-user tracking system using the extended Kalman filter (EKF).  Meanwhile, Rahal \emph{et al.} in \cite{NF3} presented a three-step algorithm for simultaneously estimating MU position and velocity, accounting for Doppler effects in the near-field.

However, the above-mentioned mobile tracking algorithms mainly rely on the signals received at the  BS, i.e., BS information. Nevertheless, the high cost and power consumption associated with radio-frequency chains often result  in a limited number of antennas at the BS. This limitation reduces the dimensionality of the BS information, thereby diminishes the localization accuracy.  However, an important yet unexplored  potential is the use of rich location-related information embedded in the signals received by the RIS, i.e., RIS information \cite{Wu5}. Due to its composition of a vast number of reflecting elements, the RIS panel can catch a much higher dimension of useful information. Actually, the idea of using RIS information for localization is perfectly matched with the newly proposed concept, the extremely large-scale RIS (XL-RIS). With an extremely large size, near-field effect exhibits and therefore each individual reflecting element on XL-RIS could collect unique AoA and ToA information. Exploiting the large number of diverse AoA and ToA patterns across the entire XL-RIS, we are able to establish a detailed and nuanced dataset and realize localization with high granularity and accuracy.

Despite the attractive benefits, the localization design based on XL-RIS information is indeed a highly challenging task  \cite{Cunhua1}. This challenge arises from the complexity of reconstructing high-dimensional XL-RIS information. On the one hand, the passive nature of XL-RIS poses a challenge in actively acquiring the signals. On the other hand, when employing the low-dimensional BS information to obtain the high-dimensional XL-RIS information, the result is not unique. This is because the same BS information corresponds to multiple distinct XL-RIS information. The uncertainty in the signal reconstruction process requires innovative algorithms to accurately capture  the XL-RIS information.

Furthermore, for mobile tracking tasks, the works in \cite{Palmucci, NF2, NF3} utilized the velocity to model the trajectory of the  MUs, employing traditional tracking algorithms like the EKF for predicting trajectory of the MU. However, traditional tracking algorithms face  challenges in complex environments, reducing localization accuracy. Specifically, the EKF linearizes MUs' nonlinear trajectories and thereby introduces the linearization error, especially exacerbated in indoor settings with multi-path and near-field effects, significantly lowering precision.  Additionally, traditional tracking algorithms inadequately leverage time-series data, particularly in recognizing long-term dependencies, restricting prediction accuracy and model adaptability.

Against the above-mentioned background, this paper investigates the mobile tracking problem with complex environments and near-field effects, and proposes an effective algorithm to enable the use of XL-RIS information. We first employ the data-driven methods. i.e., convolution neural network (CNN), to design an XL-RIS information reconstruction (XL-RIS-IR) algorithm. Building on this, we propose a comprehensive data-driven framework for MU tracking, which includes a  \emph{Feature Extraction Module} and a \emph{Mobile Tracking Module}. Our contributions are outlined as follows:

\begin{itemize}
		\item[1)] { \textbf{\emph{XL-RIS information reconstruction}}: We propose an XL-RIS-IR algorithm  for reconstructing XL-RIS information from BS information. The XL-RIS-IR algorithm includes a data processing module for preprocessing low-dimensional BS information, a DenseNet module for feature extraction, and an output module to produce the reconstructed high-dimensional RIS information.
}
		
		\item[2)] {\textbf{\emph{Comprehensive feature extraction}}: As the first component of the proposed tacking framework, we design a \emph{Feature Extraction Module} for extracting the essential features from the reconstructed XL-RIS information. This module encompasses three distinct extractors: a CNN extractor for extracting the spatial features, a time and frequency (T$\&$F) extractor aimed at extracting time and frequency domains features, and a near-field  AoAs extractor for capturing the AoAs features within XL-RIS. Finally, these features are then combined into a final feature vector for mobile tracking.  }
		
		\item[3)] {\textbf{\emph{Accurate mobile tracking algorithm}}:   The \emph{Mobile Tracking Module}, serving as the second  component of the proposed tacking framework, employs an Auto-encoder (AE) configured with a stacked bidirectional long short-term memory (Bi-LSTM) encoder and a conventional LSTM decoder. This sophisticated AE processes the   time-varying sequence derived from  the final feature vector to accurately predict the MU's three-dimensional (3D) coordinates for the upcoming time slot. }

	\item[4)] {\textbf{\emph{Confirm the effectiveness of the  framework}}:
  Simulation results reveal that the tracking accuracy of the proposed tacking framework is enhanced by utilizing the high-dimensional XL-RIS information. Moreover, the proposed  framework exhibits considerable robustness against variations in SNR. Additionally, numerical analyses confirm that our framework outperforms established benchmark models in mean squared error (MSE) metrics, highlighting its superior tracking precision.}
	\end{itemize}

\section{System Model } \label{System_Model}
Consider an XL-RIS aided uplink (UL) localization system where an  MU transmits the pilot signals to the  BS via the XL-RIS. The BS is equipped with a uniform planar array (UPA) comprising $M_1\times M_2=M$ antennas, while the MU has a single antenna. Additionally, the XL-RIS is equipped with a UPA comprising $N_1\times N_2=N$ reflecting elements.

\subsection{Geometry and Channel Model}
We now introduce the geometry configuration of the system, for the BS, the XL-RIS, and the MU. The BS is positioned at a known location \(\bm{p}_{BS} \in \mathbb{R}^3\), while the center of the XL-RIS is denoted by \(\bm{p}_{RIS} \in \mathbb{R}^3\), and the location of the \(n\)-th RIS element is represented by \(\bm{p}_{n} \in \mathbb{R}^3\) for \(1 \leq n \leq N\). The MU's location, denoted by \(\bm{p} \in \mathbb{R}^3\), is unknown and needs to be estimated.

In this system, the MU is assumed to lie in the Fresnel (radiative) near-field region of the XL-RIS. This region is defined by the distance \(d\) between the MU and the XL-RIS, which satisfies the following condition:
\begin{equation}
0.62\sqrt{\frac{D^3}{\lambda}} \leq d \leq \frac{2D^2}{\lambda},
\end{equation}
where \(\lambda\) is the carrier wavelength, and \(D\) represents the XL-RIS aperture size, defined as the largest distance between any two elements of the XL-RIS. This condition ensures that the MU is located within a specific range that allows for effective interaction with the radiative fields emitted by the XL-RIS, facilitating precise estimation of the MU's location.

Given the near-field condition, we next model the communication channel between the MU and XL-RIS. This channel is characterized by a multipath scenario, where the signal propagation can be divided into two primary components: Line-of-Sight (LoS) and Non-Line-of-Sight (NLoS).  The LoS component represents the direct propagation path from the RIS to the MU without any obstruction. Conversely, the NLoS component encompasses multiple indirect paths where the signal is reflected by various scatterers in the environment, such as walls, furniture, and human bodies.

For the LoS path from the MU to the XL-RIS, we introduce the steering vector  $\bm{a}(\bm{p}) \in\mathbb{C}^{N\times1}$,  which models the signal phase variations across the RIS elements, given the MU's position $\bm{p}$. This steering vector is derived with respect to the XL-RIS's center, denoted by $\bm{p}_{RIS}$, which is given as
\begin{equation}
[\bm{a}(\bm{p}) ]_n=\textrm{exp}\left(-j\frac{2\pi}{\lambda}(\|\bm{p}-\bm{p}_n\|-\|\bm{p}-\bm{p}_{RIS}\|)\right),
\end{equation}
for $n\in\{1,\cdots, N\}$, where  $\bm{p}_n$ represents the position of the $n$-th element on the XL-RIS.

In addition, the NLoS paths, which result from reflections of $M_s$ scatterers, are modeled through steering vectors that account for the additional path lengths. Specifically, for the $m_s$-th scatterer, the steering vector is given by
\begin{align}
[\bm{a}_{m_s}(\bm{p}) ]_n&=\textrm{exp}\bigg(-j\frac{2\pi}{\lambda} (\|\bm{p}-\bm{p}_n\| + \|\bm{q}_{m_s}-\bm{p}_n\|  \nonumber\\
&\qquad\qquad\qquad + \|\bm{p}-\bm{q}_{m_s}\| - \|\bm{p}-\bm{p}_{RIS}\| )\bigg),
\end{align}
for ${m_s}\in\{1,\cdots, M_s\}$ and $n\in\{1,\cdots, N\}$, where $\bm{q}_{m_s}$ denotes the position of the $m_s$-th scatterer.

Therefore, the overall complex channel  from the MU to the XL-RIS, incorporating both LoS and NLoS components, is thus represented as
\begin{equation}
\bm{h}(\bm{p}) = \alpha \bm{a}(\bm{p}) + \sum_{m_s=1}^{M_s} \alpha_{m_s} \bm{a}_{m_s}(\bm{p}),
\end{equation}
where $\alpha$ and $\alpha_{m_s}$ denote the complex channel attenuation coefficients for the LoS path and each NLoS path through the ${m_s}$-th scatterer, respectively.

For the communication channel between the  BS and the XL-RIS, it is assumed that no NLoS paths exist   due to the BS being situated close to the XL-RIS \cite{YjP}.  The normalized channel gain for the LoS path between BS and the XL-RIS, is represented by ${\bm G}$, whose $(m,n)$-th element is given by \cite{Bohagen1}
\begin{align}\label{4}
{\bm G}(m,n)=\textrm{exp}\bigg(j\frac{2\pi}{\lambda}r_{m,n}\bigg),
\end{align}
where $r_{m,n}$ denotes the path length between the $m$-th antenna element of the BS and the $n$-th reflecting element. Letting $\beta$ denote the common large-scale path loss attenuation, the complex channel gain of the channel between the  BS and the XL-RIS is then given by
\begin{align}\label{5}
{\bm H} =\beta{\bm G}.
\end{align}

\subsection{Signal Model}
To motivate the deployment of the XL-RIS, we assume that the LoS path between the BS and the MU is blocked, e.g., due to the buildings, cars or trees. Letting ${\bm y}_r$ denote the XL-RIS received signal, we have
\begin{align}\label{16r}
{\bm y}_r={\bm h}({\bm p}){ s},
\end{align}
where ${s}$ denotes the pilot signal transmitted by the MU. Then,  defining the phase shift vector of the XL-RIS as ${\bm \omega} = [\omega_1, \cdots, \omega_N]^T \in \mathbb{C}^{N\times1}$,   the UL signal received by the BS can be expressed as
\begin{align}\label{16}
{\bm y}={\bm H}\textrm{diag}({\bm \omega}){\bm h}({\bm p}){ s}+{\bm n},
\end{align}
where $\textrm{diag}({\bm \omega})$ and ${\bm n}$ denote the phase shift matrix of the XL-RIS and the zero-mean additive white Gaussian noise, respectively.

\section{XL-RIS Information Reconstruction (XL-RIS-IR) Algorithm}
\begin{figure*}[t]
\centering
\includegraphics[width=1\linewidth]{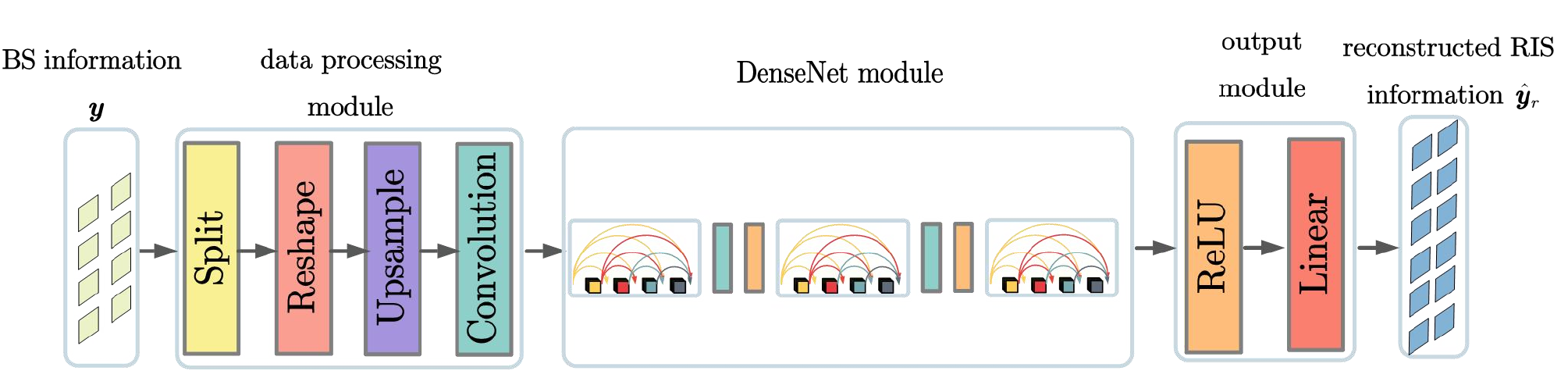}
\caption{\normalsize The structure of the XL-RIS-IR algorithm.}\label{algorithm}
\end{figure*}
 Traditionally,    localization algorithm design are based on the BS information \cite{Wu6}, e.g., ${\bm y}$. Let us further define the estimated position of $\bm{{p}}_u$ as $\bm{\hat{p}}_u$, and then the localization problem is  formulated as
  \begin{align}\label{prob}
  \underset{\bm{\hat{p}}_u }{\min} \ &|| \mathcal{H}( {\bm y})-\bm{{p}}_u ||^2,
\end{align}
where $\mathcal{H}(\cdot)$ denotes the non-linear function between \({\bm y}\)  and $\bm{\hat{p}}_u$.  The above problem can be solved with several localization algorithms which have been extensively investigated \cite{Shubo,Xing, WuFingerprint}. However, constraints such as cost and deployment conditions often limit the number of antennas, resulting in low-dimensional BS information that provides insufficient localization features, thereby potentially reducing localization accuracy. In contrast, XL-RIS  can achieve larger sizes at a lower cost, enabling them to contain more sufficient location information, which can be used to enhance localization precision. However, due to the passive nature of RISs, it is not feasible to directly obtain XL-RIS information. This necessitates the reconstruction of XL-RIS information to fully leverage its potential for improving localization accuracy.

By defining the reconstructed high-dimensional XL-RIS information as $\hat{\bm y}_r$, the RIS information reconstruction problem can be formulated as
\begin{equation} \label{pro1}
\min_{ \hat{\bm y}_r} \ \lVert {\bm y}_r - \hat{\bm y}_r\rVert^2_2,
\end{equation}
 which cannot be solved with traditional convex optimization methods, since ${\bm y}_r$ is unknown.  Hence, we design a deep learning-based XL-RIS information reconstruction (XL-RIS-IR) algorithm to solve Problem \eqref{pro1}. As shown in Fig. \ref{algorithm}, the XL-RIS-IR algorithm includes three modules: data processing module, DenseNet module, and output module, whose details are given in the following.

\subsection{Data Processing Module}
The data processing module is designed to process the input BS information $\bm{y}$. Since $\bm{y}$ is a complex vector, we  divide it into its real and imaginary parts, represented as $\mathfrak{R}({\bm y})$ and $\mathfrak{I}( {\bm y} )$, respectively. Then,  since $\bm{y}$ consists of the features of the horizontal angle domain and the vertical angle domain of the BS, we reshape the two vectors $\mathfrak{R}({\bm y})\in \mathbb{C}^{M\times 1}$ and $\mathfrak{I}( {\bm y} )\in \mathbb{C}^{M\times 1}$ into matrices, denoted as ${\bm M}_{Ry}\in \mathbb{R}^{ M_1 \times M_2}$ and ${\bm M}_{Iy}\in \mathbb{R}^{ M_1 \times M_2}$, respectively.  Then, we concatenate the reshaped matrices of the real and imaginary parts, resulting in a 3D tensor represented as
\begin{equation}\label{ty}
\boldsymbol{T}_{y} = \begin{bmatrix}
{\bm M}_{Ry}  \\
{\bm M}_{Iy}
\end{bmatrix}\in \mathbb{R}^{2 \times M_1 \times M_2}.
\end{equation}
However, the values of $M_1$ and $M_2$ may not be directly compatible with CNNs' architectures. Hence, an upsampling technique is applied:
\begin{equation} \label{tup}
\boldsymbol{T}_{\text{up}} = \text{Upsample}(\boldsymbol{T}_{y}) \in \mathbb{R}^{2 \times 256 \times 256}.
\end{equation}
The upsampling technique is employed to adjust the dimensions of the tensor $\boldsymbol{T}_{y} $  to the desired dimensions of $2 \times 256 \times 256$. It typically involves pixel value interpolation to expand the image or tensor size, redistributing pixel values accordingly. To make $\boldsymbol{T}_{\text{up}}$ suitable for processing by CNNs, which expects an input tensor of the form $(3,256,256)$, a convolutional (Con) layer is introduced to transform the bi-channel tensor to a tri-channel one:
\begin{equation}\label{tdense}
\boldsymbol{T}_{\text{Con}} = \text{Con}(\boldsymbol{T}_{\text{up}}).
\end{equation}
\begin{figure}[t!]
   \centering
		\includegraphics[width=1.0 \linewidth]{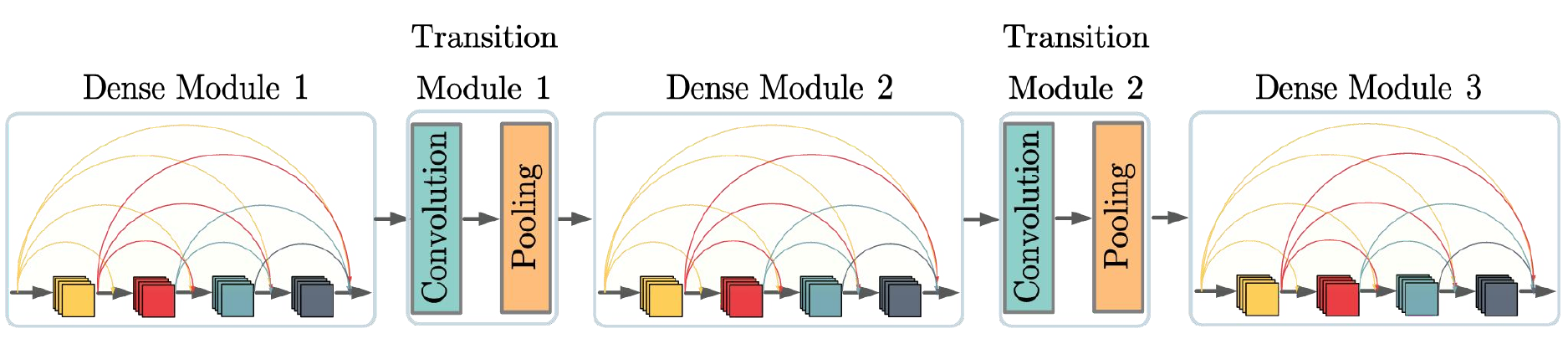}
		\caption{\normalsize Network structure of  DenseNet.}\label{DenseNet_121_Model}
\end{figure}
\begin{figure}[t]
	\centering
	\includegraphics[width=0.3\textwidth]{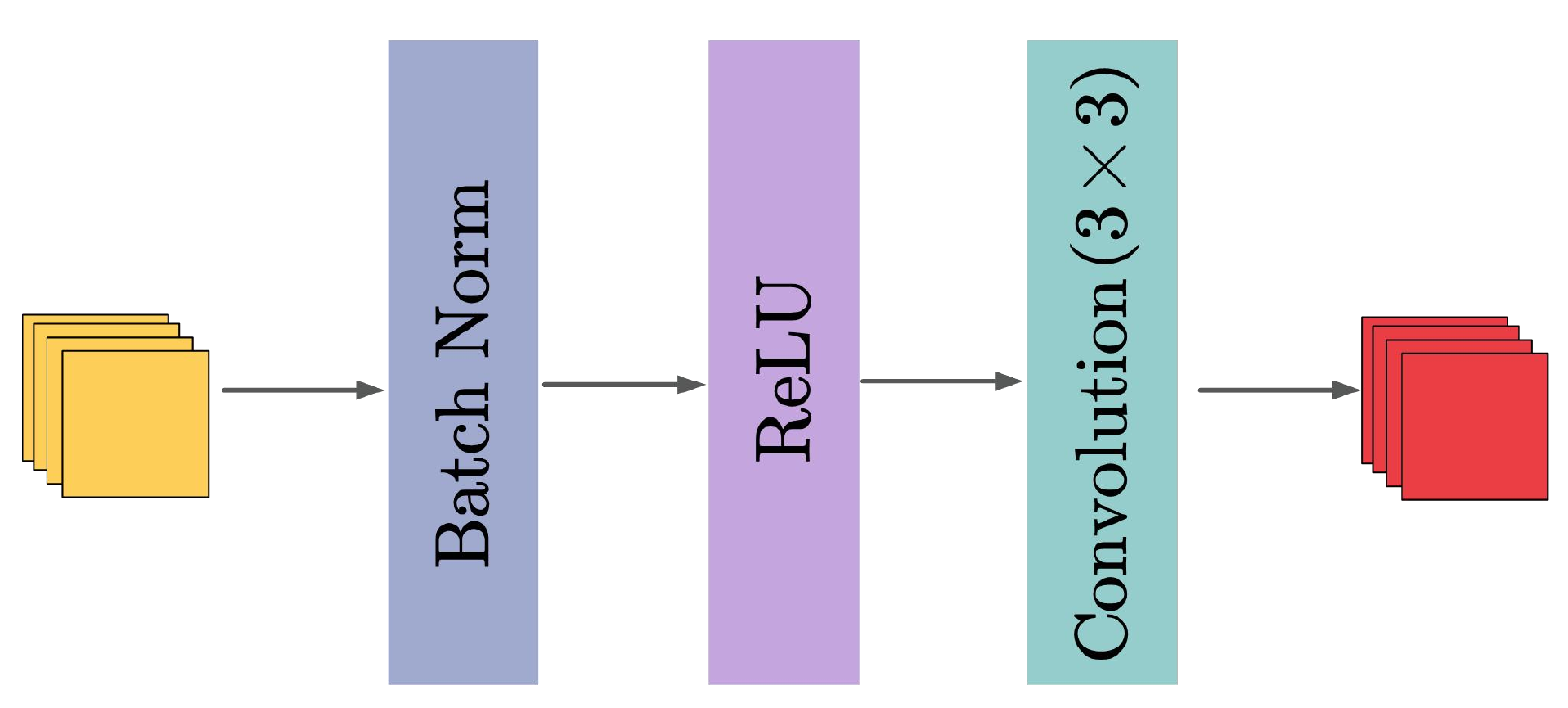}
	\caption{Network structure of a Dense Block.}
	\label{Dense_Module}
\end{figure}
\subsection{DenseNet Module}
DenseNet \cite{densenet}, inspired by the foundational principles of ResNet \cite{Resnet}, reuses the features from earlier layers. However, different from the skip connections  which might bypass a couple of layers, DenseNet establishes connections between every layer and every other layer, ensuring a more robust feature propagation. This approach, while reducing the parameter count, allows DenseNet to potentially surpass the performance of ResNet.

DenseNet   is distinguished by their excellent capability in feature extraction and is used for extracting features from the low-solution images. Besides, the intrinsic hierarchical structure of DenseNet   allows for the map from the low-level features to high-level features. Therefore, we adopt DenseNet   as a main module for reconstructing RIS information.

The design of DenseNet   is illustrated in Fig. \ref{DenseNet_121_Model}, and its architecture comprises two main components: Dense Modules and Transition Modules, which will be introduced as follows.
\subsubsection{Dense Module}
As depicted in  Fig.  \ref{DenseNet_121_Model}, the Dense Module comprises multiple Dense Blocks. The details of a Dense Block are shown in Fig.  \ref{Dense_Module}. Each Dense Block starts with a batch normalization (BN) layer, followed by a layer rectified linear unit (ReLU) activation function layer, and concludes with a $3\times3$ Con layer. A notable characteristic of the Dense Module is its connectivity: Every block incorporates the feature maps of all preceding blocks within the module.  This design ensures efficient feature reuse and a rich flow of information throughout the network \cite{densenet}.

\subsubsection{Transition Module}
From Fig. \ref{DenseNet_121_Model}, we observe that the \emph{Transition Module}, which acts as a connector between \emph{Dense Modules}, starts with a $1\times1$ Con layer, followed by a $2\times2$ average pooling layer, effectively reducing the spatial dimensions of the feature maps and streamlining them for either the next \emph{Dense Module} or the concluding output layer.

\subsection{Output Module}
After the  Dense module, we design an output module to output the reconstructed XL-RIS information. Within the output module, a ReLU layer is  used to introduce the non-linearity, enabling the network to capture and express the complex patterns of the high-dimensional RIS information. Besides,  the ReLU layer  can  address the vanishing gradient problem, facilitating better learning in deep neural networks.

Following the ReLU layer, a  {linear} layer is employed for directly outputting the desired high-dimensional XL-RIS information. This layer serves to consolidate the complex features extracted by prior layers and transforms them into a specific high-dimensional format, such as the required XL-RIS information, effectively aligning the network's output with the target data structure.
\subsection{Training of XL-RIS-IR Algorithm}
Building upon the detailed structure of the XL-RIS-IR algorithm, we initiate the direct training process. This involves inputting the received signal ${\bm y}$ into the proposed algorithm. The training aims to adjust the algorithm's parameters to achieve an optimized output, namely best reconstructing the XL-RIS information. As training progresses and converges, the algorithm efficiently transforms the low-dimensional BS information into the high-dimensional XL-RIS information.

\section{Employing XL-RIS Information For Tracking}
Once obtaining the XL-RIS information, we can use it estimate the 3D coordinates of the MU. Hence, in this section, we employ the time-varying sequence consisted by the reconstructed RIS Information, $\hat{\bm y}_r$, to determine the MU's location  when the MU is moving within a 3D indoor space.
\subsection{Tracking Problem Formulation}
The movement of the MU causes variations in the XL-RIS information, \(\hat{\bm{y}}_r\), acting as a unique fingerprint that aids in determining the MU's 3D coordinates. Therefore, in this paper, we aim  to exploit the temporal dynamics of the RIS information to accurately track the MU's movements.

To accurately document the dynamic nature of the MU's movement, we assemble a time-varying sequence that reflects the changes in \(\hat{\bm y}_r\) across a specified interval. This sequence, indexed by time slot \(t\), represents the moments when the MU transitions to a subsequent position. Formally, the sequence is expressed as \(\left[\hat{\bm y}_r(t), \hat{\bm y}_r(t+1), \ldots, \hat{\bm y}_r(t+T-1)\right]\), covering \(T\) discrete time slots. It serves to chronicle the evolution of RIS information as the MU traverses various locations.

For predicting the MU's future location at time step $T$, we consider the XL-RIS information, $\hat{\bm y}_r$, as a dynamic fingerprint that evolves over time with the MU's movement. Therefore,  we have
\begin{align}\label{aa12}
\bm{\hat{p}}_u(T) &= \mathcal{F}([\hat{\bm y}_r(t), \hat{\bm y}_r(t+1), \ldots, \hat{\bm y}_r(t+T-1)]),
\end{align}
where $\mathcal{F}(\cdot)$ is the nonlinear mapping function that transforms the time-varying sequence of $\hat{\bm y}_r$ into the estimated position, $\bm{\hat{p}}(T)$, of the MU at the subsequent time slot.  Accordingly, the mobile localization problem can be formulated as
\begin{align}
  \underset{\bm{\hat{p}}_u(T)}{\min} \ &|| \mathcal{Q}(\bm{\hat{p}}_u(T)-\bm{{p}}_u(T) ||^2.
\end{align}
aiming to minimize the discrepancy between the estimated position of the MU and its actual position at time $T$, thereby enhancing the accuracy of MU tracking.
\subsection{Tracking Framework Assisted by XL-RIS Information}
In addressing the challenges of tracking the MU, this paper introduces a comprehensive framework designed to accurately track the MU location in the near-field scenario. The proposed framework comprises two main modules: the \emph{Feature Extraction Module} and the \emph{Localization Module}.

\subsubsection{ Feature Extraction Module}
The \emph{Feature Extraction Module} is pivotal in transforming raw XL-RIS information into a meaningful representation that encapsulates spatial and temporal cues essential for localization. This module integrates three key components: a  CNN extractor, a T$\&$F feature extractor  and a near-field AoAs extractor.

\paragraph{CNN extractor} The CNN extractor is adept at capturing spatial features from the XL-RIS information by leveraging its hierarchical structure. Through Con layers, the CNN learns to identify patterns and textures in $\hat{\bm y}_r$ that are indicative of the MU's 3D coordinates, making it a powerful tool for spatial feature extraction.
\paragraph{Time $\&$ frequency (T$\&$F)   feature extractor} The T$\&$F feature extractor is designed to extract time domain and  frequency domain characteristics  to capture a comprehensive representation of the XL-RIS information.

\paragraph{Near-field AoAs extractor}  To better leverage the rich AoA information available in near-field scenarios, we have designed a specialized AoAs extractor.  This extractor is aimed at capturing AoAs features within the XL-RIS panel.

\subsubsection{Localization Module}
Within the \emph{Localization Module}, an innovative architecture featuring an Auto-encoder (AE) with a stacked bidirectional LSTM (Bi-LSTM) in the encoder and a standard LSTM in the decoder is utilized to forecast the MU's future locations.

The AE architecture employs a stacked Bi-LSTM layer within the encoder to delve into the temporal patterns and dependencies in the XL-RIS information, capturing both past and future contexts for a comprehensive understanding. The decoder, equipped with an LSTM layer, then reconstructs the future position from this encoded representation, facilitating precise prediction of the MU's location at any given time \(T\).

By integrating these two modules, our framework is specifically designed to enhance the utilization of XL-RIS information for MU tracking. Further details and insights into the XL-RIS-assisted tracking framework will be elaborated in the following two sections.
\section{XL-RIS Information Based Feature Extraction}
As the first module of the proposed tracking framework, the \emph{Feature Extraction Module} includes three kinds of  extractors, which are CNN extractor, T$\&$F extractor, AoA extractor, respectively, whose details are given as follows.
\subsection{CNN Extractor}
To efficiently handle the high-dimensional reconstructed XL-RIS information $\hat{\bm y}_r$, and extract potential location-related features,  we design a CNN-based extractor in this subsection, which includes several blocks. The details of the designed module are given as follows.
\subsubsection{Data Processing Block}
To accurately handle the preprocessing step for the reconstruction of high-dimensional  XL-RIS information $\hat{\bm y}_r$, before feature extraction via a CNN block, it is crucial to detail the preparation process of the input data. This preprocessing involves separating the XL-RIS information  into its real and imaginary components, denoted as $\mathfrak{R}(\hat{\bm y}_r)$ and $\mathfrak{I}(\hat{\bm y}_r)$, respectively. Given that $\bm{y}_r$ encapsulates features from both the horizontal and vertical angle domains of the XL-RIS, we first reshape the vectors $\mathfrak{R}({\bm y}_r) \in \mathbb{C}^{N \times 1}$ and $\mathfrak{I}({\bm y}_r) \in \mathbb{C}^{N \times 1}$ into matrices, denoted as ${\bm M}_{Ry_r} \in \mathbb{R}^{N_1 \times N_2}$ and ${\bm M}_{Iy_r} \in \mathbb{R}^{N_1 \times N_2}$, respectively. These matrices are then concatenated to form a 3D tensor represented as
\begin{equation}\label{eq:tyr}
\boldsymbol{T}_{y_r} = \begin{bmatrix}
{\bm M}_{Ry_r}  \\
{\bm M}_{Iy_r}
\end{bmatrix} \in \mathbb{R}^{2 \times N_1 \times N_2}.
\end{equation}
This 3D tensor, $\boldsymbol{T}_{y_r}$, serves as the input to the CNN, which is designed to extract location-related features from the complex XL-RIS information.

\subsubsection{CNN Block}
The CNN architecture processes the input tensor $\boldsymbol{T}_{y_r}$ through several layers, each designed to perform specific operations that contribute to feature extraction. The operation in a Con layer can be described as
\begin{align}\label{eq:CNN1}
\bm{F}_l = f_{\mathrm{ReLU}}( {\bm W}_l * \boldsymbol{T}_{y_r} + {\bm b}_l),
\end{align}
where $\bm{F}_l$ is the feature map produced by the $l$-th layer, $\bm{W}_l$ represents the weights of the filters, $\bm{b}_l$ is the bias term, $*$ denotes the convolution operation, and $f_{\mathrm{ReLU}} = \max(0, x)$ denotes the ReLU function, which is widely used as a nonlinear activation function in CNNs.

Consequently, the pooling layer is employed to reduce the dimensionality of each feature map while retaining the most important information, typically using max pooling. This procedure can be formulated as
\begin{align}\label{eq:CNN2}
\bm{P}_l = f_{\mathrm{MaxPool}}( \bm{F}_l),
\end{align}
where $\bm{P}_l$ represents the pooled feature map.

Then, the flattening layer is employed to convert the 2D feature maps into a 1D feature vector, preparing the data for processing by fully connected layers, which can be expressed as
\begin{align}\label{eq:CNN3}
\bm{g}_\textrm{flattened} = f_{\mathrm{Flatten}}( \bm{P}_L),
\end{align}
where $\bm{g}_\textrm{flattened}$ is the flattened feature vector, and $L$ is the last Con layer.

Finally, the fully connected layer is applied to allocate weights to the flattened feature vector to produce the final feature vector $\bm{T}$, encapsulating the critical information extracted from the input. This can be represented as
\begin{align}\label{eq:CNN4}
\bm{f}_\textrm{CNN} = \bm{W}_{\textrm{FC}} \cdot \bm{g}_\textrm{flattened} + \bm{b}_{\textrm{FC}},
\end{align}
where $\bm{W}_{\textrm{FC}}$ and $\bm{b}_{\textrm{FC}}$ are the weights and bias of the fully connected layer, respectively.

To provide a clear understanding of the CNN block used for feature extraction from the preprocessed XL-RIS information, the following Table \ref{tab:CNN_module} outlines each layer's definition and corresponding dimensions.

\begin{table*}[htbp]
\centering
\caption{CNN Architecture for Feature Extraction}
\label{tab:CNN_module}
\begin{tabular}{|c|c|c|}
\hline
\textbf{Layer} & \textbf{Definition} & \textbf{Dimension} \\
\hline
Input & Preprocessed Tensor & $2 \times N_1 \times N_2$ \\
\hline
Conv Layer 1 & Convolution + ReLU Activation & $N_f^{(1)} \times H_1 \times W_1$ \\
\hline
Pooling Layer 1 & Max Pooling & $N_f^{(1)} \times H_1' \times W_1'$ \\
\hline
Conv Layer 2 & Convolution + ReLU Activation & $N_f^{(2)} \times H_2 \times W_2$ \\
\hline
Pooling Layer 2 & Max Pooling & $N_f^{(2)} \times H_2' \times W_2'$ \\
\hline
Flattening Layer & Flatten Feature Maps & $N_f^{(2)} \cdot H_2' \cdot W_2'$ \\
\hline
Fully Connected Layer & Dense Layer & $N_f$ \\
\hline
\end{tabular}
\end{table*}
In this table, $N_f^{(1)}$, $N_f^{(2)}$ denote the number of filters in the first and second convolutional layers, respectively. $H_1$, $W_1$, $H_2$, $W_2$ represent the dimensions of the feature maps after each convolutional layer before pooling. $H_1'$, $W_1'$, $H_2'$, $W_2'$ are the dimensions of the feature maps after pooling. $N_f$ is the dimension of the final feature vector $\bm{f}_\textrm{CNN}$.
\subsection{T$\&$ F Feature Extractor }
In addition to the spatial features extracted by the CNN, we enhance our feature set with frequency domain characteristics and time statistical measures to capture a comprehensive representation of the XL-RIS information.  Herein, we detail the feature extraction steps, including normalization, frequency domain transformation, and time statistical analysis.
\subsubsection{Normalization}
Before feature extraction, normalization is a critical preprocessing step to ensure consistent scale across different signals, facilitating more effective analysis and comparison. Normalization adjusts the amplitude of $\hat{\bm y}_r$ to a common scale, enhancing the robustness of subsequent feature extraction steps. The process can be mathematically represented as
\begin{equation}
\hat{\bm y}_r^{\prime} = \frac{\hat{\bm y}_r - \min(\hat{\bm y}_r) \cdot \mathbf{1}}{\max(\hat{\bm y}_r) - \min(\hat{\bm y}_r)},
\end{equation}
where  $\hat{\bm y}_r^{\prime}$ is the normalized signal, $\mathbf{1}$ is a all-ones vector with the same dimensions as $\hat{\bm y}_r$, and $\min(\hat{\bm y}_r)$ and $\max(\hat{\bm y}_r)$ are the minimum and maximum values of $\hat{\bm y}_r$, respectively.

\subsubsection{Frequency Domain Transformation}
Performing the fast fourier transform (FFT) and extracting information in the frequency domain are essential for wireless localization. Firstly, the frequency domain analysis helps in identifying the presence of distinct path components in a multi-path environment, where signals from a single source reach the receiver via multiple paths due to reflections, diffractions, and scattering. Each path can have different lengths, and consequently, different phase shifts when observed in the frequency domain, providing valuable information about the relative distances and angles of arrival, which are directly related to the position of the source.  Besides, the spectral content of a signal, as revealed by the FFT, can be used to extract features that are sensitive to the positioning of the MU. For example, the Doppler shift, a change in frequency due to the relative motion between the transmitter and receiver, can be precisely identified in the frequency domain, offering insights into the velocity and direction of movement of the source relative to the MU.

 Hence, FFT is applied to $\hat{\bm y}_r^{\prime}$ to analyze its frequency components, crucial for understanding the signal's spectral properties. This procedure is formulated as
\begin{equation}
\hat{\bm y}^\text{(FFT)}_r = f_\text{FFT}(\hat{\bm y}_r^{\prime}) \in \mathbb{C}^{N_\text{yf}\times1},
\end{equation}
where $\hat{\bm y}^\text{(FFT)}_r$ denotes the frequency domain representation of the normalized signal $\hat{\bm y}_r^{\prime}$, $N_\text{yf}$ denotes frequency points of the $\hat{\bm y}^\text{(FFT)}_r$. To quantify the signal's energy distribution across the frequency spectrum, we compute the spectral energy using the squared magnitude of the FFT output, which is expressed as
\begin{equation}
\bm{e} = |\hat{\bm y}^\text{(FFT)}_r|^{\circ 2} \in \mathbb{C}^{N_\text{yf}\times1},
\end{equation}
where $(\cdot)^{\circ 2}$ denotes the element-wise square operation, transforming the complex signal into a vector of real-valued energy measurements across the frequency spectrum.

This energy vector $\bm{e}$ is then normalized to a probability distribution, which is given by
\begin{equation}
p(\hat{\bm y}^\text{(FFT)}_r) = \frac{\bm{e}}{\sum^{N_\text{yf}}_{n_\text{yf}=1} \bm{e}},
\end{equation}
allowing for the calculation of spectral entropy, which can be used as an essential feature of the reconstructed RIS information.

\subsubsection{Feature Extraction}
Several features are extracted from both the time-domain and frequency-domain representations of the signal to capture its essential characteristics.
\paragraph{Frequency Domain Features}
We compute the spectral energy and entropy as key features to capture the signal's frequency distribution. First, the spectral energy is given by
\begin{equation}
E_\text{FFT} = \sum^{N_\text{yf}}_{n_\text{yf}=1} \bm{e}.
\end{equation}
Then, the spectral entropy is formulated as
\begin{equation}
H_{FFT} = -\sum^{N_{yf}}_{n_{yf}=1}  p(\hat{\bm y}^{(FFT)}_r)  \log p(\hat{\bm y}^{(FFT)}_r).
\end{equation}

\paragraph{ Time Domain Features}
Except the frequency domain features, time domain features are also essential to the localization, especially for the MU is moving with time. To extract the time domain features, we employ the statistical time domain features,  including mean, variance. They are calculated directly from $\hat{\bm y}_r^{\prime}$, which is given as
\begin{align}
\mu = f_\text{mean}(\hat{\bm y}_r^{\prime}), \nonumber\\
\sigma^2 = f_\text{var}(\hat{\bm y}_r^{\prime}),
\end{align}
where $\mu$ is the mean, $\sigma^2$ is the variance.
\subsubsection{Feature Vector Construction}
Finally, the extracted features are concatenated to form a comprehensive feature vector $\bm{f}_\textrm{T\&F}$, which is written as
\begin{equation}
\bm{f}_\textrm{T\&F} = [\mu, \sigma^2, E_\text{FFT}, H_\text{FFT}]^T,
\end{equation}
 providing a multi-dimensional representation of $\hat{\bm y}_r$ for further localization.
\subsection{AoAs  Feature Extraction }
 In near-field localization applications, extracting AoA information within the XL-RIS panel is crucial as the near-field effects offer richer and more detailed AoA data compared to far-field scenarios.  Hence,  this subsection provide an approach to estimate the AoAs from the reconstructed XL-RIS information $\hat{\bm y}_r$.

\subsubsection{Angle Estimation}
Given the large number of different AoAs at different reflecting elements within the XL-RIS to be estimated  in near-field conditions, which increases the estimation complexity and causes signal energy to be dispersed, it is challenging to accurately estimate the AoA across the entire XL-RIS panel.  Consequently, we partition the XL-RIS into sub-arrays for angle estimation to manage the computational complexity and concentrate the signal energy within each sub-array, thus enhancing the accuracy and efficiency of AoA estimations. By doing so,  the wavefront impinging on each sub-array can be approximated as a planar wavefront due to the limited physical dimension of each sub-array, minimizing the impact of wavefront curvature. This approximation allows for the effective use of high-resolution AoA estimation techniques like MUSIC, which are predicated on the assumption of planar wavefronts, thus enabling accurate AoA estimation within each sub-array's constrained spatial domain.

Let us assume that there are  $K$ sub-arrays within the XL-RIS. To estimate $K$  AoAs using the MUSIC algorithm across $K$ sub-arrays.  For the $k$-th sub-array, where $k\in\{1,\cdots,K\}$, we have the steering vector of the $k$-th sub-array expressed as
\begin{align}
\bm{a}_k(\theta_k, \phi_k) = {\bm a}_{k}^{(e)}(\theta_{k})\otimes{\bm a}_{k}^{(a)}(\theta_{k},\phi_{k}),
\end{align}
where
\begin{align}
{\bm a}_{k}^{(e)}(\theta_k) &= \left[1, \ldots, e^{\frac{-j2\pi(N_{k,e}-1) d_x\cos\theta_k}{\lambda_c}}\right]^T, \\
{\bm a}_{k}^{(a)}(\theta_k,\phi_k) &= \left[1, \ldots, e^{\frac{-j2\pi(N_{k,a}-1) d_x\sin\theta_k\cos\phi_k}{\lambda_c}}\right]^T,
\end{align}
where $N_{k,e}$ and $N_{k,a}$ denote the numbers of elements within the $k$-th sub-array in the elevation and azimuth directions, respectively, $d_x$ denotes the distance between adjacent array elements, and $\lambda_c$ is the carrier wavelength.

The received signal at the $k$-th sub-array of the XL-RIS can be expressed as
\begin{equation}
{\bm y}_{r,k} = \alpha \bm{a}_k(\theta_k, \phi_k)s,
\end{equation}
where  $\bm{a}_k(\theta_k, \phi_k)$ denotes the array response vector. The signal $s$ corresponds to the transmitted signal. Due to the passive nature of  XL-RIS, ${\bm y}_{r,k}$ cannot be directly observed. To address this challenge, we employ the  reconstructed XL-RIS information of the $k$-th sub-array, which is denoted as $\hat{\bm y}_{r,k}$. Utilizing this information, we are able to estimate the  AoAs (e.g., $\theta_k$, $\phi_k$) across the $K$ sub-arrays. The details are given as follows.
\paragraph{Signal Covariance Matrix}
 First, by employing the FFT technique to $\hat{\bm y}_{r,k}$, we have
   \begin{align}
\hat{\bm y}^\text{(FFT)}_{r,k} &= f_\text{FFT}(\hat{\bm y}_{r,k}^{\prime}), \nonumber\\
\hat{\bm y}_{r,k}^{\prime} &= \frac{\hat{\bm y}_{r,k} - \min(\hat{\bm y}_{r,k}) \cdot \mathbf{1}}{\max(\hat{\bm y}_{r,k}) - \min(\hat{\bm y}_{r,k})},
\end{align}

Then, we  calculate the frequency covariance matrix, which is given as follows:
   \begin{align}
&\bm{R}_{f,k} \nonumber\\
&= \frac{1}{N_\text{yf}} \sum_{n_\text{yf}=1}^{N_\text{yf}} (\hat{\bm y}^\text{(FFT)}_{r,k}[n] - \bar{\bm y}^\text{(FFT)}_{r,k}) (\hat{\bm y}^\text{(FFT)}_{r,k}[n] - \bar{\bm y}^\text{(FFT)}_{r,k})^H,
\end{align}
where $\bar{\bm y}^\text{(FFT)}_{r,k}$ is the mean vector of $\hat{\bm y}^\text{(FFT)}_{r,k}$.

\paragraph{Eigen Decomposition}
For each sub-array \(k\), we perform an eigen decomposition on  \(\bm{R}_{f,k}\), which is formulated as
   \begin{align}
\bm{R}_{f,k} = \bm{E}\bm{\Lambda}\bm{E}^H,
\end{align}
where \(\bm{E}\) is a matrix whose columns are the eigenvectors of \(\bm{R}_{f,k}\), and \(\bm{\Lambda}\) is a diagonal matrix containing the eigenvalues in descending order. \(H\) denotes the conjugate transpose.

\paragraph{Identifying Signal and Noise Subspaces}
Following the results of decomposition, we proceed to determine the signal and noise subspaces, represented by \(\bm{E}_{\text{signal}}\) and \(\bm{E}_{\text{noise}}\), respectively.

\paragraph{MUSIC Spectrum}
The MUSIC spectrum for estimating $\theta_k$ and $\phi_k$ is given by the inverse of the projection of the steering vector onto \(\bm{E}_{\text{noise}}\). For each potential pair of $(\Theta_k, \Phi_k)$,   MUSIC spectrum can be derived as
 \begin{equation}
P_{\text{MUSIC},k}(\Theta_k, \Phi_k) = \frac{1}{\bm{a}_k(\Theta_k, \Phi_k)^H \bm{E}_{\text{noise},k} \bm{E}_{\text{noise},k}^H \bm{a}_k(\Theta_k, \Phi_k)}.
\end{equation}
\paragraph{Estimation of }
The AoAs, $\theta_k$ and $\phi_k$, are estimated by identifying the peaks in the MUSIC spectrum, which is given as
\begin{equation}
(\hat{\theta}_k, \hat{\phi}_k) = \arg \max_{(\Theta_k, \Phi_k)} P_{\text{MUSIC},k}(\Theta_k, \Phi_k),
\end{equation}
where $(\hat{\theta}_k, \hat{\phi}_k)$ are the estimated elevation and azimuth angles for the $k$-th sub-array.
\subsubsection{Feature Vector Construction}
Upon obtaining the AoAs estimates, $(\hat{\theta}_k, \hat{\phi}_k)$, from the MUSIC algorithm for each of the $K$ sub-arrays, we aggregate these estimations into a structured feature set.   The collective feature set $\mathcal{F}$, representing the AoA estimates across all sub-arrays, can be constructed as:
\begin{equation}
\mathcal{F}_\textrm{AoA} = \left\{(\hat{\theta}_1, \hat{\phi}_1), (\hat{\theta}_2, \hat{\phi}_2), \ldots, (\hat{\theta}_K, \hat{\phi}_K)\right\}.
\end{equation}
Alternatively, for numerical and analytical convenience, these AoA estimates can be organized into a vector form $\bm{f}_\textrm{AoA}$, which is given as
\begin{equation}
\bm{f}_\textrm{AoA}  =\left[\hat{\theta}_1, \hat{\phi}_1, \hat{\theta}_2, \hat{\phi}_2,\cdots, \hat{\theta}_K, \hat{\phi}_K   \right].
\end{equation}
\subsection{Final Feature Vector Construction}
After combing the feature vector ${\bm f}_\text{CNN}$,  the T$\&$F feature vector   ${\bm f}_\text{T$\&$F}$ and the AoAs feature vector ${\bm f}_\text{AoA}$, we have the  final feature vector, denoted as ${\bm f}_\textrm{Final}$,  which is given by
\begin{equation}
\bm{f}_\text{Final} = \left[{\bm f}_\text{CNN}, {\bm f}_\text{T$\&$F}, {\bm f}_\text{AoA}\right],
\end{equation}
which will be fed into the following \emph{Mobile Tracking Module}.
\section{XL-RIS Information Based Tracking Algorithm}
In this section, we employ the final feature vector ${\bm f}_\textrm{Final}$  for tracking the MU using the data-driven DL based framework. Data-driven DL based framework is now widely designed and employed for MU localization and channel estimation.  However, all these DL-based localization   work ignore the time sequence of the recieved signal when the MU is moving on a specific trajectory. In another word, the adjacent observations of the time-varying recieved signal can be further utilized for more precise estimation.

There are two widely employed techniques called recurrent neural network (RNN) and long short term memory (LSTM) on solving the time-varying tasks, such as natural language processing. Both techniques consider the information from the previous entered data and the currently entered data to estimate. Specifically, RNN has feedback loops to maintain information over time. However, it is difficult for RNN on learning long-term temporal dependencies due to the vanishing gradient problem. Differently, LSTM introduces input and forget gates for better preservation of long-term dependencies on dealing gradient flow.

Therefore, in this section,  we aim to develop an Autoencoder (AE) equipped with a stacked bidirectional long short-term memory (Bi-LSTM) encoder and a standard LSTM decoder, referred to as the Stacked Bi-LSTM AE algorithm. This design is intended to accurately predict the 3D coordinates of the mobile user (MU) in the subsequent time slot.
\subsection{LSTM}
Before presenting the details of the Stacked Bi-LSTM AE algorithm, the details of  LSTM network are given as follows.

The LSTM  network, an enhancement over traditional RNN, introduces a gradient-based learning algorithm capable of connecting past information to the current task. LSTMs process time-sequence data in a chain-like structure, moving in a forward direction. The unique aspect of LSTM architecture compared to standard RNNs is its hidden layers within each LSTM cell. The details of  LSTM network are given as follows.
\subsubsection{Forget Gate}
The first layer of the LSTM network is called forget gate  which is also named forget layer,  which is formulated as
  \begin{equation}
    {\bm f}_{t} = \sigma\left( {\bm W}_f {\bm x}(t) + \tilde{\bm W}_f\tilde{\bm x}(t-1) + {\bm b}_f\right),
    \end{equation}
where $\sigma$ denotes the sigmoid activation function, $\tilde{\bm x}(t-1)$ denotes the information passed from previous layer, ${\bm x}(t)$ represents  the input feature vector at time step $t$, while $\tilde{\bm W}_f$, ${\bm W}_f$ and ${\bm b}_f$ denote the weight and bias of forget gate, respectively.
\subsubsection{Input Gate}
Consequently, the second layer is called input layer which is also named input gate,  which can be  expressed as
 \begin{equation}
    {\bm i}_{ t} = \sigma\left( {\bm W}_i{\bm x}(t) + \tilde{\bm W}_i\tilde{\bm x}(t-1) + {\bm b}_i\right),
 \end{equation}
 where $\tilde{\bm W}_i$, ${\bm W}_i$ and ${\bm b}_i$ denote the weight and bias of input gate, respectively.
 \subsubsection{Cell Input State}
Then, to update the cell input state, e.g., ${\bm c}_{t}$, $ \tanh$ function is employed and calculated as
\begin{equation}
   {\bm c}_{t} = \tanh\left( {\bm W}_c {\bm x}(t) + \tilde{\bm W}_c\tilde{\bm x}(t-1) + {\bm b}_c\right).
    \end{equation}
where $\tilde{\bm W}_c$, ${\bm W}_c$ and ${\bm b}_c$ denote the weight and bias of cell input state, respectively.
\subsubsection{Output Gate}
To decide the next hidden state, the output gate ${\bm o}_{t}$ is utilized, which is written as
    \begin{equation}
    {\bm o}_{t} = \sigma\left( {\bm W}_o {\bm x}(t) + \tilde{\bm W}_o\tilde{\bm x}(t-1)  +{\bm b}_o\right),
    \end{equation}
where $\tilde{\bm W}_o$, ${\bm W}_o$ and ${\bm b}_o$ denote the weight and bias of output layer, respectively.
\subsubsection{Output State}
Apart from the cell input state ${\bm c}_{g,t}$ in the hidden layers, there are two other cell states in the structure: the previous cell output state ${\bm c}(t-1)$ fed in the current LSTM cell, and the current cell output state ${\bm c}(t)$ passed to the next LSTM cell. The output state at the current time $t$ can be formulated  as
\begin{equation}
{\bm c}(t) = {\bm f}_{t} \odot {\bm c}(t - 1) + {\bm i}_{t} \odot {\bm c}_{t}.
\end{equation}
where $\odot$ denotes the Kronecker product.
\subsubsection{Output Layer}
The conventional input layer is  $ {\bm x}(t)$ at the time slot $t$. Finally, the output layer can be formulated as
\begin{equation}
\tilde{\bm x}(t) =  {\bm o}_{t} \odot \tanh( {\bm c}(t)).
\end{equation}
\subsection{ Stacked Bi-LSTM}
To accurately capture the dynamic nature of the input time-varying feature sequence, the Bi-LSTM is employed when constructing encoder. This design choice is pivotal because it allows the model to maintain the temporal continuity and dependencies between each time step in the sequence. By integrating both forward and backward LSTM cells, the architecture ensures that information from both past and future contexts is utilized, facilitating a more comprehensive understanding of the temporal patterns. This bidirectional approach is especially beneficial for tasks like mobile localization tracking, which recognizing the sequence's temporal evolution.

Furthermore, it has been proved that by stacking multiple hierarchical models, the performance can be improved progressively. Hence, we adopt a stacked structure where the output from the lower layer is then fed as the input to the upper layer with $L_s\geq2$ Bi-LSTM layers. The workflow of the stacked Bi-LSTM considers both forward and backward directions and deeper structure with $T$ time slots. During $T$ time slots, we have the time sequence given by ${\bm F}_\textrm{fea}=[{\bm f}_\textrm{Final}(t),{\bm f}_\textrm{Final}(t+1), \cdots, {\bm f}_\textrm{Final}(t+T)]$. By employing this time varying sequence from the past $T$ time slots, the location of the MU at $T+1$ time slot can be estimated via the design AEBi-LSTM. The details of the AE Bi-LSTM are given as follows subsections.
\subsection{Encoder Construction for Stacked Bi-LSTM AE}
In this subsection, we present  the details of the encoder constructed by the stacked Bi-LSTM, which is designed to extract the time-varying sequence ${\bm F}_\textrm{fea}$.
\subsubsection{The first Bi-LSTM Layer}
In a Bi-LSTM layer, the input time-varying  feature sequence  is processed to capture both past (forward) and future (backward) dependencies within the sequence. The first Bi-LSTM is employed for estimating the location of the MU at the $T+1$ time slot  by using the time-varying sequence ${\bm F}_\textrm{fea}$,  whose details are given as follows.
\paragraph{Forward Pass in Bi-LSTM}
The forward pass processes the input sequence from time $t$ to time $t+T$, capturing information as it moves forward in time.
\paragraph{Backward Pass in Bi-LSTM}
The backward pass processes the sequence in reverse, from time $t+T$ to time $t$, capturing future context for each time step. The operations are analogous to those of the forward pass but applied in reverse order.
\subsubsection{Stacked $L_s$ Bi-LSTM Layers}
After introducing the first Bi-LSTM layer, the construction of the Encoder in an AE Bi-LSTM setup involves stacking additional Bi-LSTM layers to further refine and process the features. This stacked structure enables the model to capture more complex temporal dependencies in the sequence data, enhancing its predictive capabilities for tasks like estimating the MU location based on historical data.

Following the first Bi-LSTM layer, we incorporate $\mathrm{L}_s$ Bi-LSTM layers. Each subsequent layer receives the output from all time slots of its predecessor as input. Consequently, each Bi-LSTM layer builds upon the patterns and dependencies identified in the time-varying sequence by the layer before it, extracting increasingly abstract features from the sequence.

Within each Bi-LSTM layer, temporal steps are processed akin to the first layer, employing forward and backward LSTM units to assimilate information from past and future contexts, respectively. Thus, the model's output at each time step synthesizes current input features with contextual insights drawn from both preceding and succeeding steps.

To mitigate overfitting, Dropout layers can be interspersed between Bi-LSTM layers. These Dropout layers randomly nullify a fraction of the neurons' activations during training, thereby enhancing the model's generalization ability.

The output generated by the last Bi-LSTM layer in the Encoder is typically used to represent a compressed and high-level feature representation of the entire input sequence. This representation captures the core information and patterns of the input sequence, providing necessary context for the Decoder.

\begin{figure*}
    \centering
        \begin{subfigure}{.30\textwidth}
        \centering
        \includegraphics[width=1\linewidth]{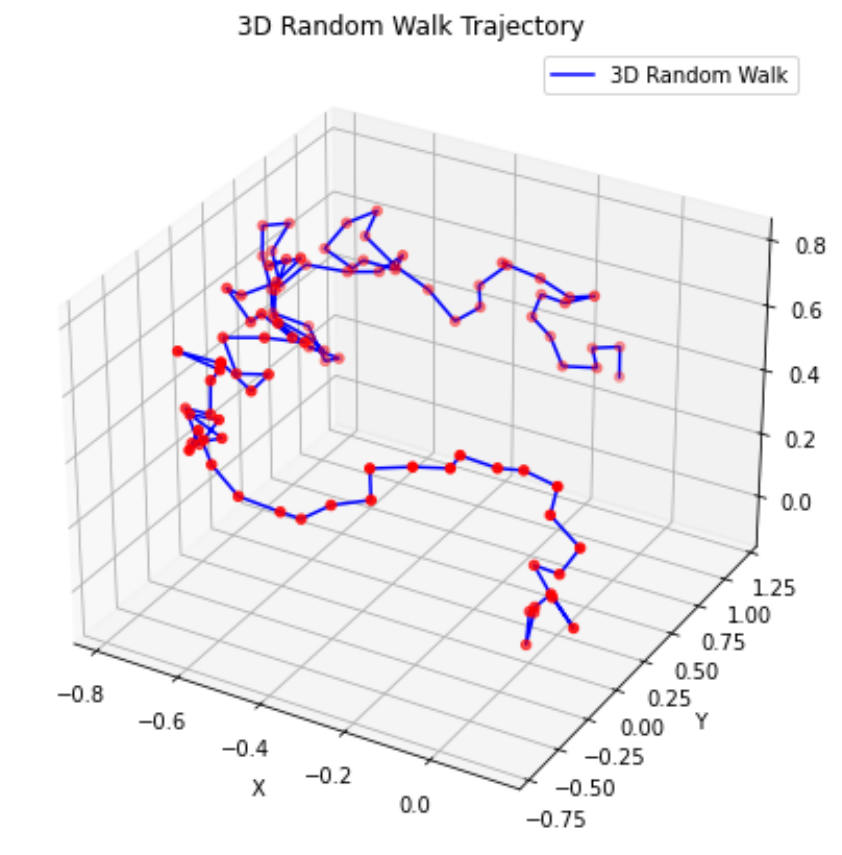}
        \caption{3D random walk example.}
        \label{3D_random}
    \end{subfigure}%
        \hfill
     \begin{subfigure}{.30\textwidth}
        \centering
        \includegraphics[width=1\linewidth]{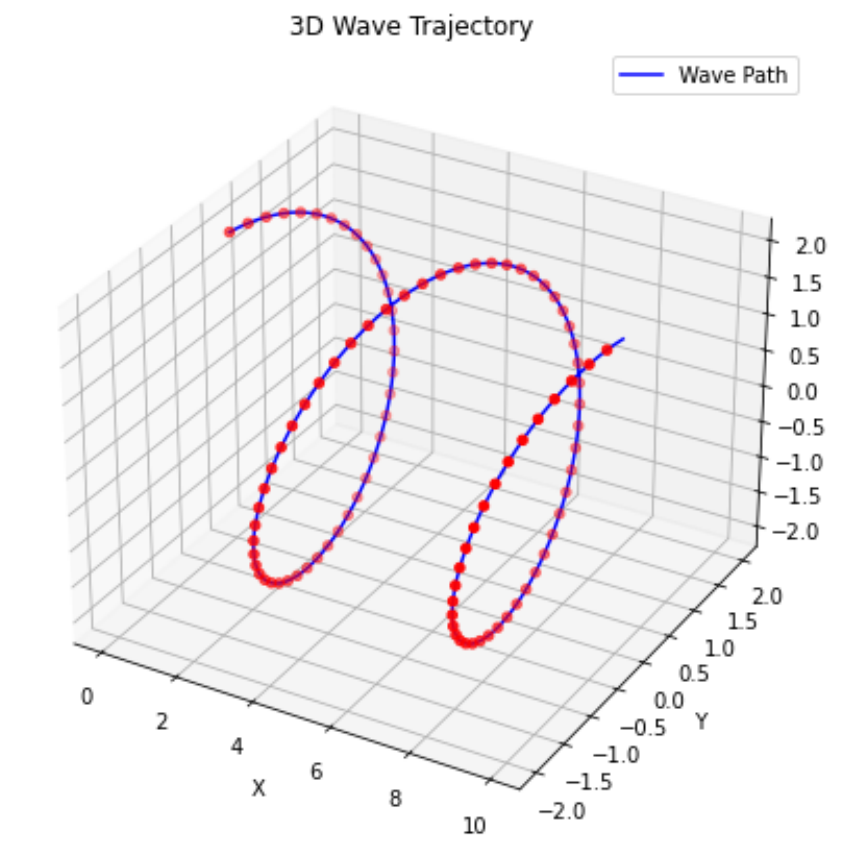}
        \caption{3D wave example.}
        \label{3D_wave}
    \end{subfigure}
    \hfill
    \begin{subfigure}{.30\textwidth}
        \centering
        \includegraphics[width=1\linewidth]{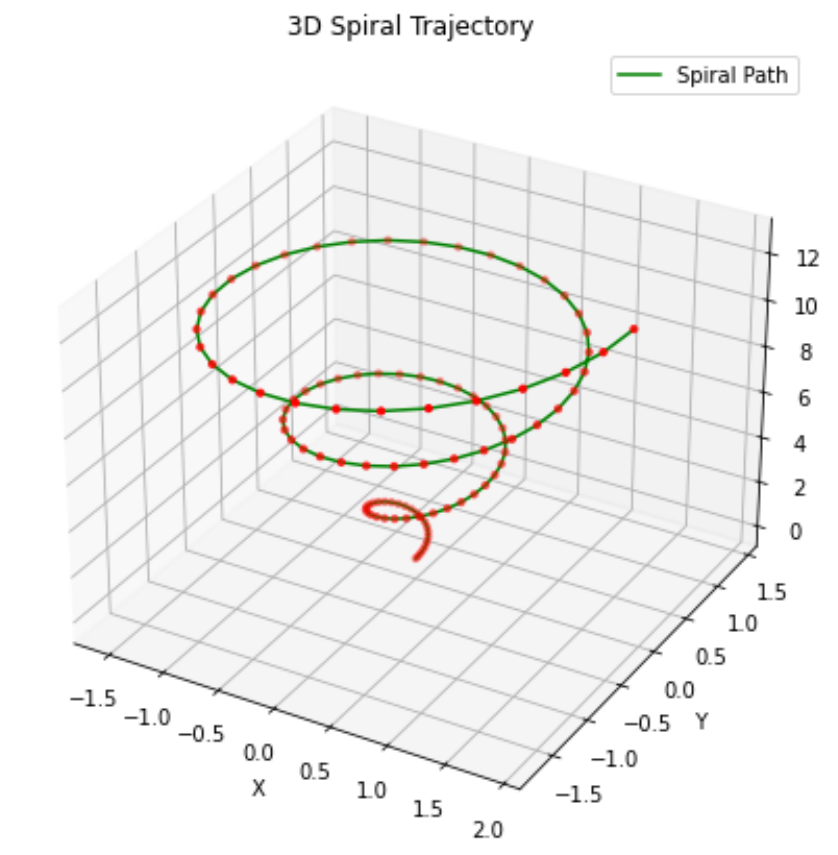}
        \caption{3D spiral example.}
        \label{3D_spiral}
    \end{subfigure}%
 \hfill
    \caption{Examples of different types of trajectory. }\label{trajectories}
\end{figure*}
\begin{figure*}
    \centering
        \begin{subfigure}{.30\textwidth}
        \centering
        \includegraphics[width=1\linewidth]{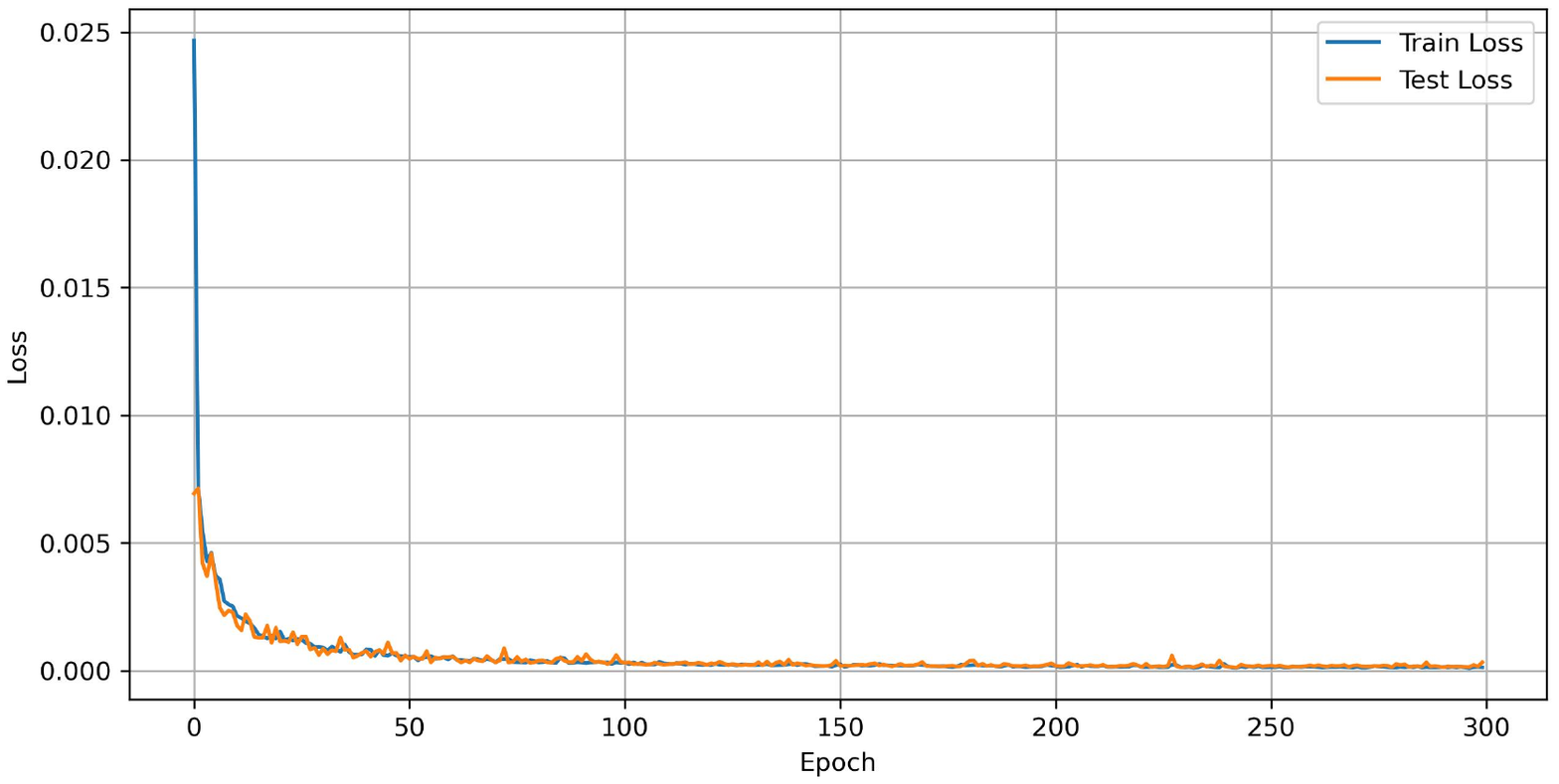}
        \caption{Test and train losses of random walk trajectory.}
        \label{random_loss}
    \end{subfigure}%
        \hfill
     \begin{subfigure}{.30\textwidth}
        \centering
        \includegraphics[width=1\linewidth]{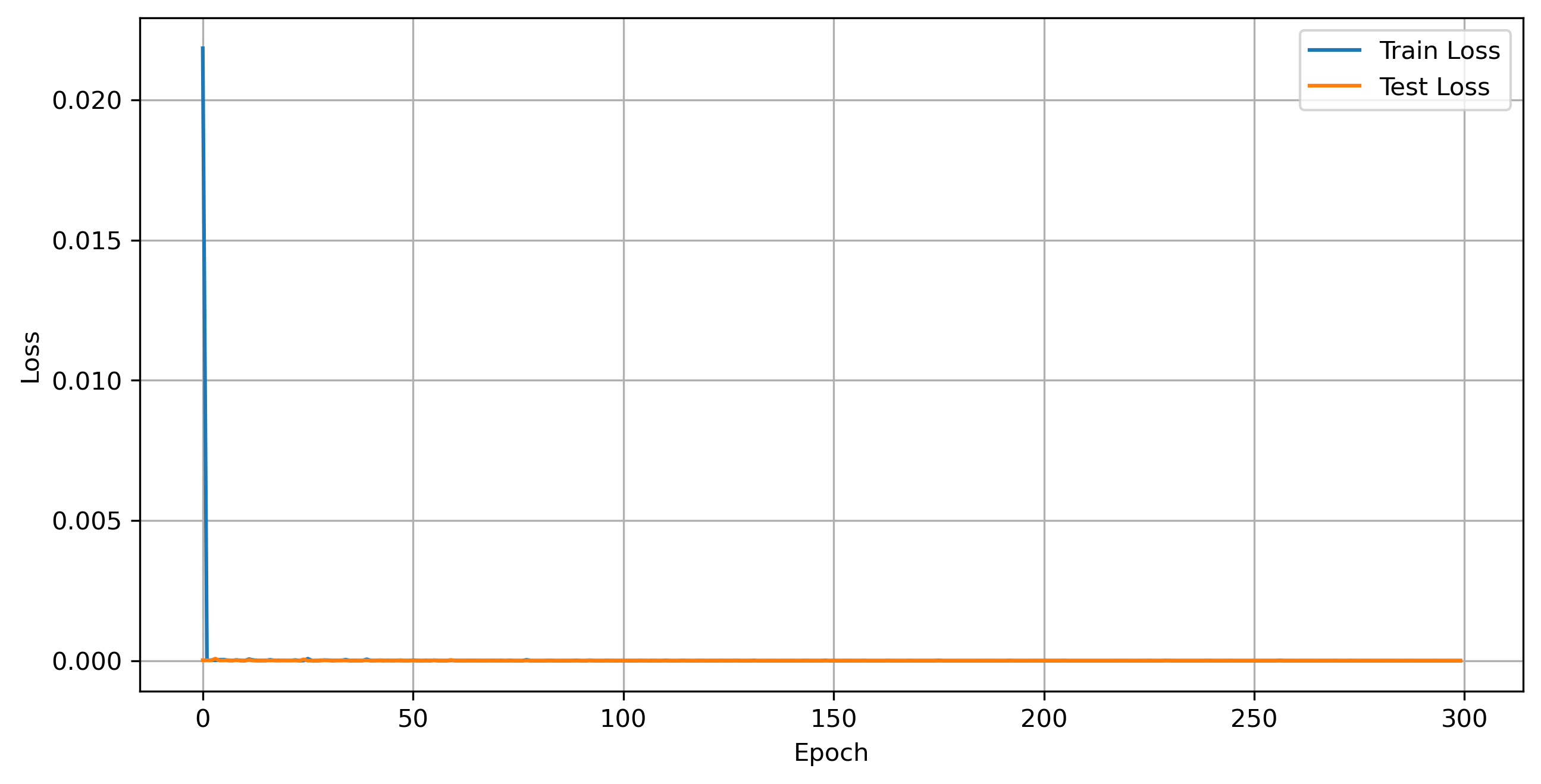}
        \caption{Test and train losses loss of  spiral trajectory.}
        \label{spiral_loss}
    \end{subfigure}
    \hfill
    \begin{subfigure}{.30\textwidth}
        \centering
        \includegraphics[width=1\linewidth]{Spiral_loss}
        \caption{Test and train losses of  wave trajectory.}
        \label{wave_Loss}
    \end{subfigure}%
 \hfill
    \caption{Losses of different types of trajectory.}\label{Loss}
\end{figure*}
\subsection{Decoder Construction for Stacked Bi-LSTM AE}
After constructing the encoder of the Stacked Bi-LSTM AE, the next step is to build the decoder to predict the location of the MU at the $T+1$ time slot. The decoder utilizes the high-level features generated by the encoder to perform this prediction.
\subsubsection{Initialization}
The encoder processes the entire input sequence and abstracts the input through its hierarchy into a high-level representation that captures the essence of the input data, including timing dependencies and patterns. Hence, the final hidden state generated by the encoder is employed for the initial hidden state of the decoder. By initializing the decoder with this state, we provide a strong starting point for the generation process, ensuring that the decoder has all the context needed to generate accurate and relevant output.

\subsubsection{Layers Construction for Decoder}
The decoder consists of one LSTM layers, one dropout layer, one dense layer, where the details of each layers are given as follows.
\paragraph{LSTM Layer}
In the context of predicting the location at time $T+1$, a single LSTM layer is chosen for the decoder instead of a  Bi-LSTM or a Stacked Bi-LSTM configuration due to the nature of the prediction task. Bi-LSTM layers, which process information in both forward and backward directions, are more suited to situations where the entire sequence is known and the context from future steps is as important as the past, which is not the case when predicting future locations based solely on past and current information. Stacked LSTMs, while powerful for capturing complex patterns in more intricate sequences, increase model complexity and computational cost, which might not be necessary for tasks where the encoder has already provided a condensed and meaningful representation of the input sequence. Thus, a single LSTM layer is efficient and sufficient for leveraging the encoded context to predict the next location, minimizing the risk of overfitting while maintaining model simplicity and interpretability.
\paragraph{Dropout Layer}
Dropout is included as a regularization technique. By randomly omitting portions of the layer's inputs, dropout forces the network to learn more robust features that are useful in conjunction with many different random subsets of the other neurons. This reduces the likelihood of overfitting, especially when training data is limited or when the architecture is complex relative to the amount of training data.
\paragraph{Dense Layer}
Dense layer is used to produce a specific output size from the LSTM layer's output, matching the dimensionality of the prediction task (e.g., 3D coordinates for location). The linear activation  in the output Dense layer allows for the direct prediction of continuous values.  The linear activation  in the output Dense layer allows for the direct prediction of continuous values.

\section{Simulation Results}
In this simulation, we consider a setup in a 3D space, involving an MU, an RIS, and a BS.  The RIS is composed of $ N_1 \times N_2 = 10 \times 10 = 100$ elements. The center of the RIS is located at $\bm{p}_r=(6, 0, 2)$ m. The BS comprises $ M_1 \times M_2 = 4 \times 4 = 16$ antennas. The center of the BS is positioned at $\bm{p}_b=(0, 5, 1.5) $ m. The MU-RIS link includes $ P = 10 $ propagation paths. The transmit power of the MU is $30$ dBm.

To simulate the movement trajectories of objects, we have defined three types of paths: random walk, wave, and spiral, to model the possible trajectories that objects may follow in reality.  Examples of these three distinct trajectories are illustrated in Fig. \ref{trajectories}. These trajectories are generated with a random starting point within a space limited to \(10 \times 10 \times 3\) $\text{m}^3$. Over \(10000\) such paths are generated, each comprising \(11\) steps. For each step along a trajectory, we generate the signal received at both the BS and RIS, denoted by ${\bm y}_t$ and ${\bm y}_r$, respectively.    Besides, the principles behind the generation of the three types of trajectories are as follows:
\begin{enumerate}
   \item \textbf{Random Walk Trajectory:}  We start by selecting a random starting point within the indoor space. For each step in the trajectory, we generate a unit vector pointing in a random direction within 3D space, and move along this unit vector's direction by a step size randomly chosen from the range  $(0,1]$.    An example of random walk is given in Fig. \ref{3D_random}.
   \item  \textbf{Wave Trajectory:} To generate a wave trajectory, we start by picking a random point. We then draw this path along the x-axis, setting its length and dividing it into steps. Then, based on the wave's amplitude of $2$  and wavelength of $5$,  a  sine function is employed to modify  the y-coordinates. This creates the familiar peaks and troughs of a wave. Throughout, we keep the z-coordinate constant, ensuring the wave moves in a two-dimensional plane. Fig. \ref{3D_wave} presents an example of the  wave trajectory.

  \item  \textbf{Spiral Trajectory:} We create a spiral path by using the formula $r = a + b\theta$. Here, $r$ is how far each point is from the random start point, $a=0.1$ is the starting distance from the center, and $b=3$ decides how much this distance increases.  $\theta$ increases step by step, making the spiral trajectory. By changing from polar coordinates (using $r$ and $\theta$) to regular 3D coordinates, we can draw this spiral in a 3D space. Fig. \ref{3D_spiral} presents an example of the generated spiral trajectory.

\end{enumerate}

 \begin{figure}[t]
\centering
\includegraphics[width=1\linewidth]{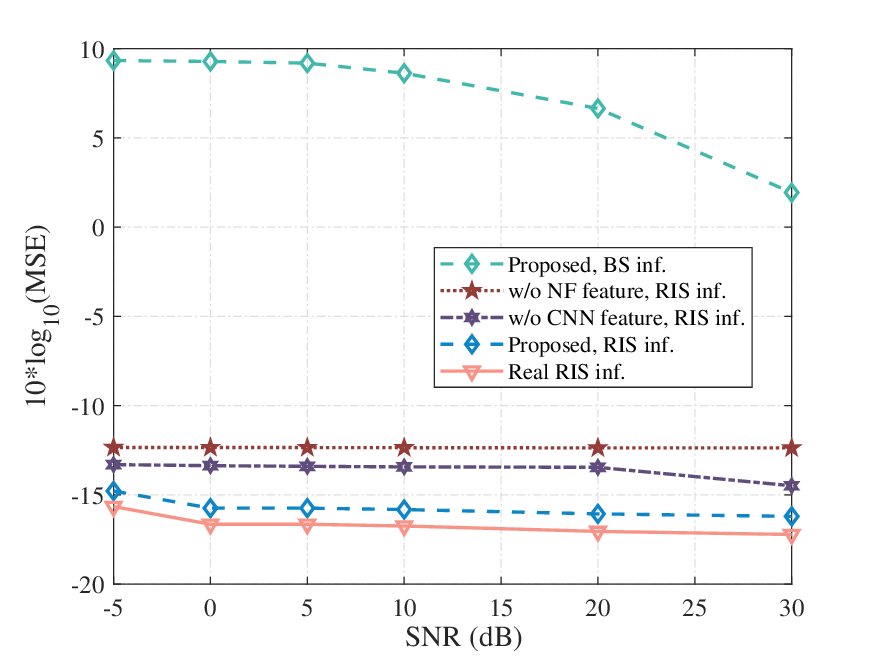}
\caption{The MSE comparison of different inputs for proposed algorithm  versus SNR (dB).}\label{NF_com}
\end{figure}

\begin{figure}[t]
\centering
\includegraphics[width=1\linewidth]{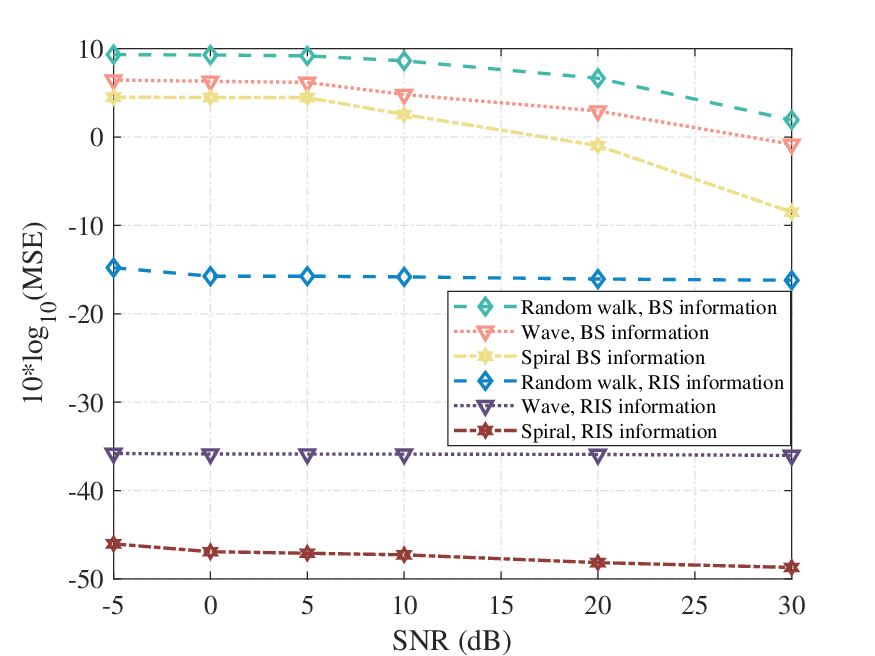}
\caption{The MSE comparison of different types of trajectory  versus SNR (dB).}\label{NF_tra}
\end{figure}

Fig. \ref{Loss} showcases the training and test loss curves of three different kinds of trajectories. Specifically, Fig. \ref{random_loss} displays the iterative convergence behaviors of training and test  losses for the random walk trajectory. It is evident that by approximately $100$ epochs, both training and test losses for the random walk trajectory have converged. Although there are minor fluctuations in the test loss afterwards, their magnitude is negligible. Furthermore, Fig. \ref{spiral_loss} and Fig. \ref{wave_Loss} depict the convergence trends of training and testing losses for the spiral and wave trajectories, respectively. For both trajectory types, the training and testing losses converge by the $15$-th epoch, exhibiting minimal fluctuations thereafter. As we can see from Fig. \ref{Loss}, due to the regular patterns of the spiral and wave trajectories, the proposed algorithm is capable of quickly learning their movement patterns. In contrast, the random walk, characterized by its unpredictable nature, requires more epochs for the neural network to learn. However, the eventual convergence of the random walk as well demonstrates the effectiveness of the proposed algorithm.

Fig. \ref{NF_com} shows the mean squared error (MSE) comparison among three scenarios: without near-field feature extraction, without  CNN feature extraction, and with feature extraction. Fig. \ref{NF_com} reveals that utilizing near-field feature extraction alone yields better results than relying solely on CNN feature extraction. This underscores the necessity of extracting near-field features, suggesting that CNN may not fully capture all relevant localization information on its own. Besides, as shown in Fig. \ref{NF_com}, compared to the curve without near feature extraction, incorporating our near-field feature extraction significantly enhances performance, highlighting the value of tailored near-field  feature extraction in improving localization accuracy.

 Fig. \ref{NF_com} also displays the MSE performance curve for localization using actual RIS information, labeled as `Real RIS inf.'. The closeness of this curve to the `Proposed, RIS inf.' curve, which represents localization using reconstructed RIS information, validates the effectiveness of our proposed algorithm for XL-RIS information reconstruction. It shows that the localization accuracy achieved with our reconstructed information closely approximates that achieved with the actual RIS information, supporting the reliability of our reconstruction approach in mimicking real RIS information for localization purposes.

Fig. \ref{NF_tra} presents the comparison of  MSE for different trajectories versus the SNR in dB. As depicted in Fig. \ref{NF_tra}, the MSE values associated with both the wave and spiral trajectories are significantly lower than those of the random walk trajectory. This distinction arises because the wave and spiral trajectories exhibit structured movement patterns, so that the deployed Stacked Bi-LSTM AE  network can analyze more effectively. Besides, it is also noteworthy that utilizing the received signal at the BS, e.g., BS information, results in substantially poorer performance compared to the algorithms when XL-RIS information is used.  This further validates the efficacy of employing high-dimensional RIS  information for near-field  tracking, as proposed in our study. Moreover, as shown in Fig. \ref{NF_tra}, when employing the high-dimensional RIS  information,  the  accuracy maintains a certain level of accuracy even under low SNR conditions with a random walk trajectory, which demonstrates robustness of proposed algorithms against low SNR.
\begin{figure}[t]
\centering
\includegraphics[width=1\linewidth]{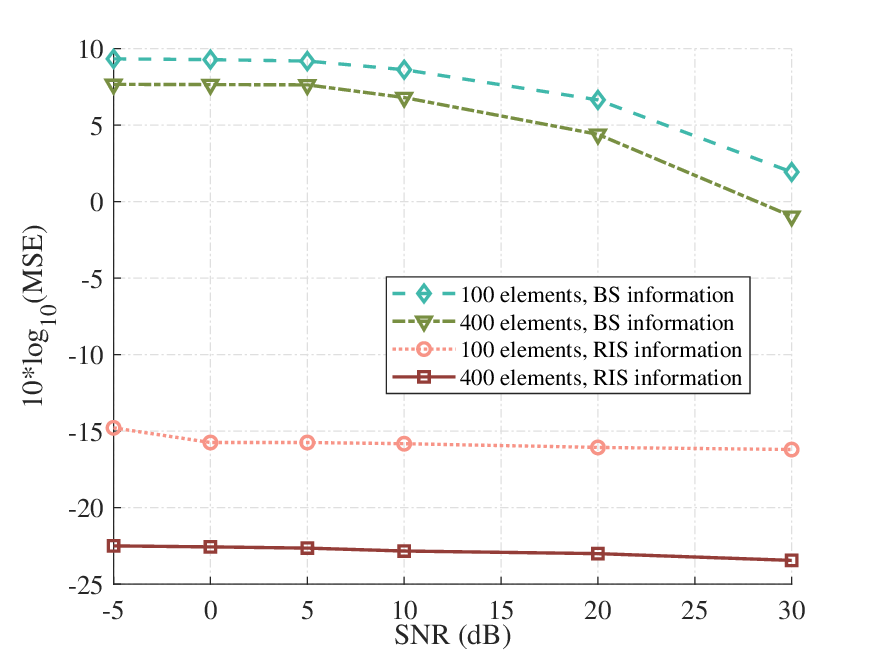}
\caption{The MSE comparison of different number of reflecting elements within the RIS  versus SNR (dB).}\label{NF_num}
\end{figure}

As shown Fig. \ref{NF_num}, it is evident that increasing the number of reflective elements in the RIS panel enhances positioning accuracy, for both cases employing BS information and RIS information. However, it can be observed that increasing the number of reflecting elements from  $100$ to $400$, the improvement of localization accuracy when using BS information is not significantly pronounced. Besides,  it necessitates a higher  SNR  to achieve better localization results. In contrast, when employing RIS information, as shown in Fig. \ref{NF_num}, the accuracy with $400$ reflective elements is significantly higher than that with $100$. This emphasizes the importance of using XL-RIS information for improved positioning accuracy in near-field scenarios.

\begin{figure}[t]
\centering
\includegraphics[width=1\linewidth]{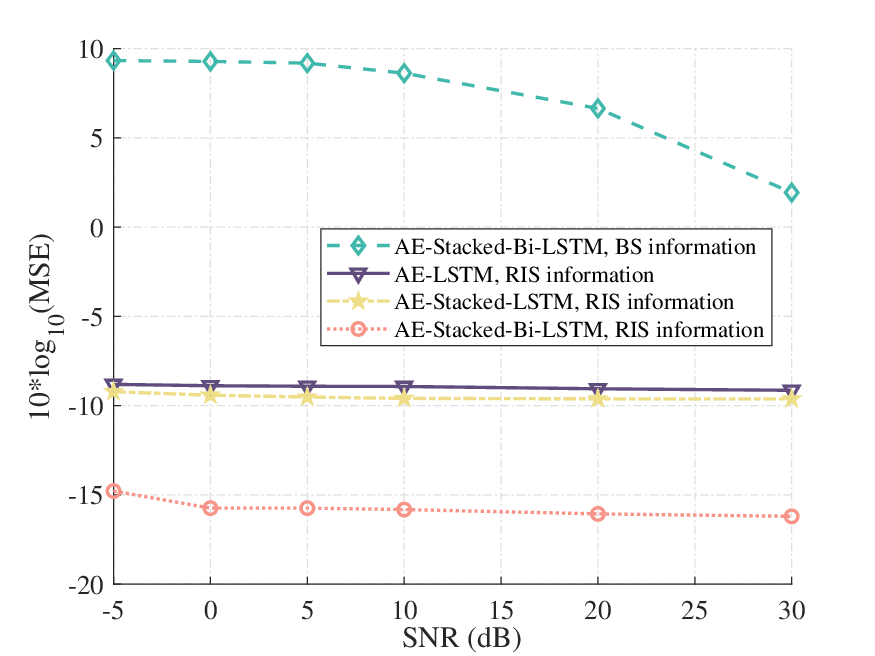}
\caption{MSE performance across SNR (dB) for localization algorithms with and without feature extraction.}\label{NF_algorithm}
\end{figure}

 Fig. \ref{NF_algorithm} showcases a comparison of MSE across SNR for different algorithms. As shown in Fig.  \ref{NF_algorithm}, `AE-Stacked-LSTM' refers to an architecture that incorporates multiple layers (stacked) but does not employ bidirectional processing, while `AE-LSTM' utilizes a simpler, single-layer structure without the stacking or bidirectional enhancements. The observed results clearly demonstrate that the proposed algorithm, referred to as `AE-Stacked-Bi-LSTM', significantly surpass the other algorithms in localization performance. This superiority suggests that the bidirectional processing, when combined with a stacked approach, effectively captures the temporal dependencies and spatial characteristics of the RIS information, thus enhancing the predictive accuracy and proving the proposed methodology's efficacy.

\section{Conclusion}\label{Con}
This paper has introduced a mobile tracking framework leveraging the high-dimensional signal received from  XL-RIS information. We have presented  an XL-RIS-IR algorithm to reconstruct the high-dimensional  XL-RIS information from the low-dimensional BS information.  Building on this, we have  proposed  a comprehensive framework for mobile tracking, consisting of a \emph{Feature Extraction Module} and a \emph{Mobile Tracking Module}.  Simulation results have demonstrated that the tracking accuracy of our proposed framework is enhanced by the   reconstructed XL-RIS information. Besides, it has been validated that the proposed framework exhibited the robustness to SNR variation.

\bibliographystyle{IEEEtran}
				\bibliography{myre}

			\end{document}